\documentclass[prb,preprint]{revtex4} 

\usepackage{amsmath} 
\usepackage{graphicx}

\newcommand{\Ev}{\mathbf{E}}
\newcommand{\Bv}{\mathbf{B}}
\newcommand{\xv}{\mathbf{x}}
\newcommand{\dv}{\mathbf{d}}
\newcommand{\jv}{\mathbf{j}}
\newcommand{\rv}{\mathbf{r}}
\newcommand{\lv}{\boldsymbol{\ell}}

\newcommand{\tvhat}{\hat{\mathbf{t}}}
\newcommand{\nvhat}{\hat{\mathbf{n}}}
\newcommand{\Nvhat}{\hat{\mathbf{N}}}

\newcommand{\deltav}{\boldsymbol{\delta}}
\newcommand{\grad}{\boldsymbol{\nabla}}
\newcommand{\muv}{\boldsymbol{\mu}}
\newcommand{\xiv}{\boldsymbol{\xi}}

\begin{document}

\title{Magnetic dipoles and electric currents}

\author{Guido Corb{\`o}}
\author{Massimo Testa}

\affiliation{Dipartimento di Fisica, Universit\`a di Roma ``La
Sapienza", Sezione INFN di Roma\\ P.le A. Moro 2, 00185 Roma, Italy}

%\date{\today}

\begin{abstract}
We discuss several similarities and differences between the
concepts of electric and magnetic dipoles. We then consider the
relation between the magnetic dipole and a current loop and show
that in the limit of a pointlike circuit, their magnetic fields
coincide. The presentation is accessible to undergraduate
students with a knowledge of the basic ideas of classical
electromagnetism.
\end{abstract}

\maketitle

The concept of a magnetic dipole describes the long distance limit of the field
produced by a steady current flowing in a small loop of wire.\cite{Feynman,Jack,Blinder,Tellegen,Rao} 
The word ``dipole'' is borrowed from electrostatics but when
used in magnetostatics, this terminology is somewhat deceptive
because a magnetic dipole is physically very different from its
electric counterpart.
The aim of this paper is to discuss the similarities and differences
of these concepts.

Recall the definition of an electric dipole. We
start with a configuration in which two charges $+q$ and $-q$
($q>0$) are located at $\deltav/2$ and $-
\deltav/2$ respectively. The electric dipole
is obtained by taking the limit $\delta\rightarrow0$
keeping fixed the quantity
\begin{equation} \label{1}
\dv\equiv q
\deltav,
\end{equation}
which is called the {\em electric dipole
moment}.
The dipole electric field $\Ev_d$ can be obtained from the potential \cite{unit}
\begin{equation} \label{2}
V_d(\xv)=-\frac{1}{4\pi}\dv\cdot
\grad \frac{1}{|\xv|},
\end{equation}
so that
\begin{equation} \label{3}
\Ev_d(\xv)=-\grad V_d(\xv)= \grad \Big(\frac{1}{4\pi}\dv \cdot
\grad \frac{1}{|\xv|}\Big).
\end{equation} 

It might be tempting to define a magnetic dipole with moment
$\muv$ in a similar way: that is, the object
which generates the magnetic field
\begin{equation} \label{4}
\Bv_d(\xv)= \grad\Big(\frac{1}{4\pi} \muv\cdot
\grad \frac{1}{|\xv|} \Big).
\end{equation} 
However, Eq.~(\ref{4}) is inconsistent with the nonexistence of magnetic
monopoles, as described by the Maxwell equation
\begin{equation} \label{due}
\grad \cdot \Bv=0,
\end{equation}
because
\begin{equation} \label{5}
\grad \cdot \Bv_d(\xv) =
\frac{1}{4\pi}\muv \cdot \grad
\Big(\grad\frac{1}{|\xv|} \Big)=-
\muv\cdot
\grad\delta^{(3)}(\xv)\neq 0,
\end{equation}
In Eq.~\eqref{5} we used the result\cite{distr}\begin{equation} \label{6}
\nabla^2\frac{1}{|\xv|}=-4\pi
\delta^{(3)}(\xv).
\end{equation} 

The failure to satisfy Eq.~(\ref{due}) is not surprising because $\Bv_d$ in Eq.~\eqref{4} was constructed
as the limit of zero separation between monopole and
anti-monopole, which in the magnetic case do not exist.

A modification of Eq.~(\ref{4}) at the origin \cite{Casimir,Griff}
\begin{equation} \label{7}
\Bv_d(\xv)= \grad \Big(\frac{1}{4\pi} \muv\cdot
\grad \frac{1}{|\xv|} \Big)+
\muv\delta^{(3)}(\xv)
\end{equation}
fixes the
problem and gives a divergenceless field. However, the field given by Eq.~(\ref{7}) is no longer
conservative (irrotational), in contrast to its electric counterpart, Eq.~(\ref{3}).

The difference between electric and magnetic fields is that, in a stationary situation, the electric field is conservative as a
consequence of the Faraday equation \begin{equation} \label{new2}
\grad\times \Ev=-
\frac{1}{c}\frac{\partial \Bv}{\partial t}=0,
\end{equation} 
whereas the magnetic field, which is divergenceless, cannot also be irrotational (unless it is identically zero).

In a world without monopoles, a magnetic dipole
must be defined in terms of current distributions only. The magnetic
effects of a steady current density $\jv$ are
described by Ampere's equation 
\begin{equation} \label{uno}
\grad\times \Bv=\frac{\jv}{c} .
\end{equation} 
From Eq.~\eqref{uno} we can calculate the magnetic field
$\Bv$ provided the condition,
\begin{equation} 
\label{uno2} \grad\cdot \jv =0,
\end{equation} 
which is equivalent to conservation of charge in the steady case, is
satisfied. Equation~(\ref{uno}) shows that the non-conservative part of the
magnetic field is located at the points at which the current
density is nonzero.

Therefore in the magnetic dipole case, Eq.~(\ref{7}), the only
contribution needed to satisfy Ampere's equation is the term
proportional to
$\muv\delta^{(3)}(\xv)$ because
\begin{equation} 
\grad\times \Bv_d
= \grad\times[\muv\delta^{(3)}(\xv)]=
-\muv\times \grad\delta^{(3)}(\xv).
\end{equation} 
We shall now show that $\Bv_d$ given by Eq.~(\ref{7})
is the magnetic field generated by a current loop of
infinitesimal size.

We start from the solution of Eqs.~(\ref{due}) and (\ref{uno}) which
can be found in textbooks on electromagnetism:
\cite{Feynman,Jack}
\begin{equation} \label{dieci}
\Bv(\xv)=\frac{1}{4\pi c}
\grad\times \!\int\! d^3\xi\,
\frac{\jv(\xiv)}{r},
\end{equation}
where $r=|\xv-\xiv|$
is the
distance between the generic point $\xiv$ of the
integration region and the observation point $\xv$.
For a coil $\gamma$ made of a thin wire, Eq.~(\ref{dieci})
becomes\cite{Feynman}
\begin{equation} \label{dieci2}
\Bv_\gamma(\xv)=\frac{i}{4\pi c}
\grad\times \!\oint d\ell\,
\frac{\tvhat}{r}=\frac{i}{4\pi c}\!\oint d\ell\,
\frac{\tvhat \times \rv}{r^3},
\end{equation} 
where the line integral, with length element $d\ell$, runs over the
wire whose tangent unit vector is denoted by
$\tvhat$.
The circuit $\gamma$ in Eq.~(\ref{dieci2}) must be closed because of
Eq.~(\ref{uno2}), and $i$ is the (constant) current
in the circuit.

\begin{figure}[h!] 
\begin{center} 
\includegraphics{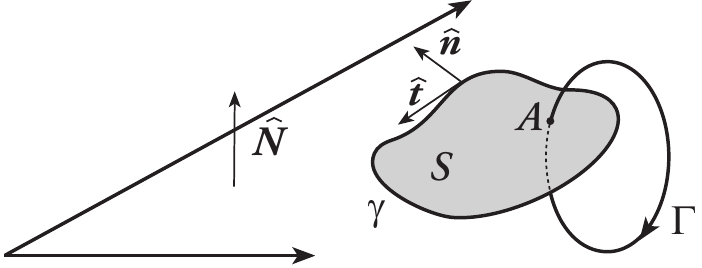}
\caption{\label{Massimo_TestaFig01.pdf}The circuit $\gamma$ and the closed path $\Gamma$ used to evaluate
the circulation of $\Bv_\gamma$.} 
\end{center} 
\end{figure}

We assume that the current loop is a plane circuit enclosing an
area $S$. We denote by $\Nvhat$ the unit
vector orthogonal to the plane, oriented according to the right-hand
rule with respect to $\tvhat$. We also denote
by $\nvhat$ the external normal to the wire
(see Fig.~\ref{Massimo_TestaFig01.pdf}). These unit vectors are related by
\begin{equation} \label{ventiquattro}
\tvhat = \Nvhat\times \nvhat.
\end{equation}
If we substitute Eq.~(\ref{ventiquattro}) into Eq.~(\ref{dieci2}), we
obtain 
\begin{equation} \label{venticinque}
\Bv_\gamma
(\xv)=\frac{i}{4\pi c}\oint d\ell\, \frac{(\Nvhat \times \nvhat)
\times \rv}{r^3}.
\end{equation} 
We use the identity
\begin{equation} 
\label{ventisette}\frac{(\Nvhat\times \nvhat)
\times \rv}{r^3}=-
(\Nvhat\times \nvhat)
\times \grad_x\frac{1}{r}=
-\nvhat\Big(\Nvhat\cdot \grad_x\frac{1}{r})
+ \Nvhat(\nvhat\cdot \grad_x\frac{1}{r} \Big),
\end{equation}
and write Eq.~(\ref{venticinque}) as
\begin{equation} 
\label{venticinque.2}
\Bv_\gamma
(\xv)=\frac{i}{4\pi c}\!\oint d\ell
\left[-\nvhat \Big(\Nvhat\cdot \grad_x\frac{1}{r} \Big)
+ \Nvhat \Big(\nvhat\cdot \grad_x\frac{1}{r} \Big)\right].
\end{equation} 
If we use Green's formula in two dimensions \begin{equation} 
\label{venticinque.3}
\oint\!f \nvhat\, d\ell=\!\int_S
\grad_\xi f\,d\sigma,
\end{equation}
where $d\sigma$ is the
surface element of $S$, and the relation
\begin{equation} \label{venticinque.4}
\grad_x\frac{1}{r}=-\grad_\xi\frac{1}{r},
\end{equation}
we obtain
\begin{subequations}
\begin{align}
\label{trenta}
\Bv_\gamma (\xv) & = \frac{i}{4\pi c}
\grad_x \left[\!\int_S
(\Nvhat\cdot \grad_x\frac{1}{r})\,d\sigma
\right] + \Nvhat \frac{i}{c} \int_S
\delta^3(\xv-\xiv)\,d\sigma \\
&
\equiv \Bv^{(1)}_\gamma
(\xv)+\Bv^{(2)}_\gamma
(\xv).
\end{align}
\end{subequations}
The gradient $\Bv^{(1)}_\gamma$ is irrotational
and is nonzero in all of space, in contrast to
$\Bv^{(2)}_\gamma$ which is non-zero only inside the
plane region $S$ delimited by the coil $\gamma$. 

It is instructive to show how $\Bv_\gamma$
satisfies Ampere's law in its integral form, that is, \begin{equation} 
\label{trenta21} \oint_{\Gamma}\Bv_\gamma\cdot
d\lv =\frac{i}{c},
\end{equation} 
where $\Gamma$ is any closed path
linked with $\gamma$ as shown in Fig~\ref{Massimo_TestaFig01.pdf}. Because $\Bv^{(1)}_\gamma$ is a pure gradient, we
have 
\begin{equation}
\label{trenta3} \oint_{\Gamma}\Bv_\gamma\cdot
d\lv=
\oint_{\Gamma}\Bv^{(2)}_\gamma\cdot
d\lv.
\end{equation} 
The integral on
the right-hand side of Eq.~(\ref{trenta3}), by virtue of the delta function,
has a contribution only from the
point of intersection $A$ between $\Gamma$ and $S$, which leads to Eq.~(\ref{trenta21}).
Equation~(\ref{trenta3}) is surprising because it shows that Ampere's law
is satisfied only by $\Bv^{(2)}_\gamma$, which is the part of the magnetic field localized
inside $\gamma$.

To make contact with the dipole field $\Bv_d$
given by Eq.~(\ref{7}), we take the limit as the coil area goes to zero, keeping the product $\mu \equiv iS/c$ constant. We
have
\begin{subequations}
\begin{align} \label{trentotto} \Bv^{(1)}_\gamma
(\xv) & = \frac{i}{4\pi c} \grad_x
\left[ \int_S
(\Nvhat\cdot \grad_x\frac{1}{r})\,d\sigma
\right] \\
& = \frac{\mu}{4\pi} \grad_x \left[\frac{1}{S}
\int_S
(\Nvhat\cdot \grad_x\frac{1}{r})\,d\sigma
\right] \\
& = \frac{\mu}{4\pi} \grad_x \overline{\left[
(\Nvhat\cdot \grad_x\frac{1}{r})
\right]}, \label{last}
\end{align}
\end{subequations}
where the bar in Eq.~\eqref{last} denotes the mean value in $S$. By the
mean value theorem we know that
\begin{equation} \label{trentanove}
\overline{\left[
(\Nvhat\cdot \grad_x\frac{1}{r})
\right]}= \left[
(\Nvhat\cdot \grad_x\frac{1}{r})
\right]_{\overline{P}}
\end{equation}
where $\overline{P}$ is a suitable point
inside $S$. In the limit of pointlike $S$, we have
\begin{equation} \label{quaranta}
\Bv^{(1)}_\gamma
(\xv) = \frac{1}{4\pi} \grad_x \left[({\muv}\cdot \grad_x\frac{1}{r})
\right]
\end{equation}
where
\begin{equation} \label{trentasette} \muv =
\frac{iS}{c} \Nvhat
\end{equation}
may be identified with
the magnetic moment of the small loop and $r$ is the distance
between the observation point and the position $\xiv_0$
of the (pointlike) circuit.

As for $\Bv^{(2)}_\gamma (\xv)$, which
contains a delta function, the pointlike limit must be discussed using generalized functions.\cite{distr}
We introduce a test function
$f(\xv)$, which is an infinitely differentiable function
vanishing at infinity faster than any inverse power of
$|\xv|$,\cite{distr2}, and study the $S
\rightarrow 0$ limit of expressions such as 
$\int\Bv^{(2)}_\gamma (\xv)
f(\xv)\,d^3x$.
From Eq.~(\ref{trenta}) we have
\begin{subequations}
\begin{align} \label{quarantuno}
\int\Bv^{(2)}_\gamma
(\xv) f(\xv)\,d^3x & = 
\Nvhat \,\frac{i}{c} \int d^3x\,
f(\xv)\int_S
\delta^3(\xv-\xiv)\,d\sigma
\\
& = \Nvhat \,\frac{i}{c} \int_S
f(\xiv)\,d\sigma \\
& = \muv \frac{1}{S}\int_S
f(\xiv)\,d\sigma\rightarrow
\muv f(\xiv_0),
\end{align}
\end{subequations}
when $S$ shrinks to the point $\xiv_0$. This implies
\begin{equation} \label{quarantadue}
\Bv^{(2)}_\gamma
(\xv) \rightarrow
\muv\delta(\xv-\xiv_0).
\end{equation} 
If we compare with Eq.~(\ref{7}), we find that an infinitesimal current
loop generates a magnetic field identical to the one given by a
magnetic dipole of moment
$\muv =(iS/c) \Nvhat$.

\begin{acknowledgements}
We thank the reviewers for valuable suggestions on how to improve the presentation of our paper.
\end{acknowledgements}

\section*{Figure caption}

\begin{thebibliography}{5}

\bibitem{Feynman}R. P. Feynman, R. B. Leighton, and M. Sands, \textsl{The
Feynman Lectures on Physics} (Addison-Wesley, Reading,
MA, 1999), Vol. 2.

\bibitem{Jack} J. D. Jackson, \textsl {Classical Electrodynamics} (John Wiley
$\&$ Sons, New York, 1998), 3rd ed.

\bibitem{Blinder} S. M. Blinder, ``Delta functions in spherical coordinates and how to avoid losing them: Fields of point charges and dipoles,'' Am. J. Phys. {\bf 71}, 816--818 (2003).

\bibitem{Tellegen} B. D. H. Tellegen, ``Magnetic-dipole models,'' Am. J. Phys. {\bf 30}, 650--652 (1962).

\bibitem{Rao} N. D. Rao, ``A note on the vector potential of a magnetic dipole,'' Am. J. Phys. {\bf 39}, 1276--1277 (1971).

\bibitem{unit} We use rationalized cgs units.

\bibitem{distr} I. M. Gel'fand and G. E. Shilov, {\em Generalized
Functions} (Academic Press, New York, 1964), Vol. 1.

\bibitem{Richards} J. I. Richards and H. K. Youn {\em The Theory of Distributions A Nontechnical Introduction} (Cambridge University Press, 1995)

\bibitem{Casimir} H. B. G. Casimir, {\em On the Interaction Between the Atomic Nuclei and Electrons} (W. H. Freeman, San Francisco, 1963). 

\bibitem{Griff} D. J. Griffiths, ``Hyperfine splitting in the ground state of hydrogen,'' Am. J. Phys. {\bf 50}, 698--703 (1982).

\bibitem{distr2} More precisely, we deal with tempered distributions. \cite{Richards}



\end{thebibliography}
\end{document}